\documentclass[reprint,aps,pra,showpacs,floatfix,nofootinbib]{revtex4-1}
\usepackage{graphicx,amsmath,amssymb,amsthm}
\usepackage[bookmarks=true]{hyperref}

\newcommand{\ii}{\mathrm{i}}
\newcommand{\e}{\mathrm{e}}
\newcommand{\dd}{\mathrm{d}}

\newcommand{\cU}{\mathcal{U}}
\newcommand{\cV}{\mathcal{V}}
\newcommand{\cE}{\mathcal{E}}

\newcommand{\Span}{\mathrm{span}}
\newcommand{\ket}[1]{|#1\rangle}
\newcommand{\bra}[1]{\langle #1 |}
\newcommand{\ketbra}[2]{|#1\rangle\!\langle #2|}

\begin{document}
\title{Gate-teleportation-based blind quantum computation}
\author{Mear M. R. Koochakie} 
\email{koochakie@gmail.com}
\affiliation{Department of Physics, Sharif University of Technology, Tehran 14588, Iran}


\begin{abstract}
Blind quantum computation (BQC) is a model in which a computation is performed on a server by a client such that the server is kept blind about the input, the algorithm, and the output of the computation.
Here we layout a general framework for BQC which, unlike the previous BQC models, does not constructed on specific computational model.
A main ingredient of our construction is gate teleportation.
 We demonstrate that our framework can be straightforwardly implemented on circuit-based models as well as measurement-based models of quantum computation.
 We illustrate our construction by showing that universal BQC is possible on correlation-space measurement-based quantum computation models.
\end{abstract}
\maketitle
\section{Introduction}
From Shor's factoring algorithm \cite{ShorAlg-94} to the BB84 quantum key distribution protocol \cite{BB84}, quantum physics has enabled to surpass classical computational algorithms and communication protocols.
A recent application of quantum physics in the privacy of delegated computations has emerged in the blind quantum computation protocols \cite{SecAssQC-Childs-QIC05,UBQC-Broadbent-IEEE09}.
A blind computation is a protocol which allows a client to compute on a remote server with full privacy about the computation and the results. Hence, the client delegates a computation to the server without letting the server know about the input, the computation, and the output.
It seems difficult (if not impossible) to perform universal classical computation on a server blindly using only a classical communication channel and classical computers \cite{HidingInf-Abadi-JCSS89}. 
However, with a quantum channel and a quantum computer in the server's lab blind computation becomes possible \cite{SecAssQC-Childs-QIC05,UBQC-Broadbent-IEEE09}.
A proof-of-principle experimental implementation of this idea 
has also been preformed recently \cite{UBQCexp-Barz-Sci12}.
A number of other universal BQC scenarios has also been proposed, which mainly differ with each other in implementation and efficiency \cite{BQCWeek-Dunjko-PRL12,CVBQC-Morimae-PRL12,AncDrivBQC-Sueki-PRA13,EfficientUBQC-Giovannetti-PRL13,OptimalBQC-Mantri-PRL13,AKLTBQC-Moriame-QIC15}.

As far as we know, all of the proposed BQC scenarios are based on specific models. 
For example, the model proposed in Ref.~\cite{UBQC-Broadbent-IEEE09} is based on one-way measurement-based quantum computation \cite{OneWay-Rauss-PRL01}, whereas the model of  Ref.~\cite{EfficientUBQC-Giovannetti-PRL13} is more suited for programmable quantum gate arrays \cite{PQGA-Nielsen-PRL97}. 
In this paper, we lift this model dependence, and propose a fairly general framework for BQC which can be more conducive for direct implementation on various computational models such as general measurement-based or circuit-based quantum computation, obviating the need for simulating the framework from one model to another.
Our framework requires the server to be able to perform Bell-basis measurements on one of its internal qubits and the qubits sent by the client, and next can swap these qubits. 
In addition,  the client is required to be able to prepare a family of entangled two-qubit states which are locally equivalent to a Bell pair.

\section{Privacy of BQC protocols}
Let $\mathcal{Q}$ be a subset of $\mathsf{BQP}$ (bounded error quantum polynomial time computational complexity class).
Let $P$ be a quantum computation protocol in which two parties are involved, 
which should collaborate to solve a problem, 
one as ``client'' and the other as ``server,'' sharing a classical and a quantum channel with each other. 
We assume the client has less quantum power than the server 
in the sense that the client alone cannot solve all instances of the problems in $\mathcal{Q}$ efficiently, but the server alone has quantum devices that can solve any problem in $\mathcal{Q}$ efficiently.
We call $P$ a blind quantum computation protocol on $\mathcal{Q}$ if the client can solve any problem in $\mathcal{Q}$  without revealing either of the problem, the algorithm, and the answer to the server. 

The description of any computation is transferred to the server by using both classical and quantum channels.
The states in the quantum alphabet should not be mutually orthogonal; otherwise, the quantum channel basically acts like a classical channel.
The security of the whole protocol is now related to how much information the server can obtain from the channels, and what methods to use to prevent the server from such intrusions. 

Placing ``traps'' in the computation is perhaps a straightforward method to detect a privacy-invasive and non-cooperative server.
In the course of computation, the client instructs the server to perform a given set of randomly chosen  non-entangling operations on a few qubits using the BQC protocol---
these qubits are chosen by the client (without the knowledge of the server) as the traps. 
Next the client randomly checks the states of the trap qubits.
Since there is no method to discriminate non-orthogonal states  perfectly, 
if the server tries to find any clue about the computation from the quantum communications, there is always a nonzero chance he would make mistake about the operation to be performed on the trap qubits.
Hence, a periodic direct or indirect checking of the traps would catch the cheating server.

There also exists another privacy ensuring method for BQC protocols, which employs a secret hidden in the classical instructions sent by the client to the server that effectively erases the information inside the quantum communications about the computation \cite{UBQC-Broadbent-IEEE09,CompSecDeleQC-Dunjko-arXiv14}.
Later in one of our examples, we elaborate on a similar privacy mechanism for BQC---section~\ref{sec:ubqc-without-trap}.

\section{Basic idea}
In our construction of BQC, we shall employ the idea of gate teleportation (GT). Thus we call our scheme ``gate-teleportation-based blind quantum computation'' (GTBQC).
Here, we shortly review the idea of gate teleportation~\cite{PQGA-Nielsen-PRL97,GateTele-Gottesman-Nature99,StoringQD-Vidal-PRL02}.

Consider the standard quantum teleportation protocol applied on the state $\ket{\psi}$. 
Now rather than using a Bell pair, $\ket{\Phi^+}:= \tfrac{1}{\sqrt{2}}(\ket{00}+\ket{11})$, let us use the state 
\begin{equation}
\ket{\Phi_V}:=\openone \otimes V \ket{\Phi^+},
\end{equation}
in which $V$ is an arbitrary single-qubit unitary operation.
Thus the state of the teleported qubit becomes
\begin{equation}
V \sigma \ket{\psi},
\end{equation}
where $\sigma \in \Sigma:=\{\openone,X,Y,Z\}$ (Pauli matrices plus identity) is the by-product of the Bell-basis measurement of the teleportation process. 

We will employ this basic gate-teleportation idea later throughout our construction in order to store a single-qubit operation in an entangled state and apply it later on demand.
Consider $\cV=\{V_i\}$ as the set of unitaries from which the client makes the states $\ket{\Phi_{V_i}}$.
For the protocol to be blind, at least two of the states $\{\ket{\Phi_{V_i}}\}$ need to be non-orthogonal.
That is, $
\langle \Phi_{V_i}|\Phi_{V_j} \rangle =\frac{1}{2} \mathrm{tr}\, V_i^\dagger V_j,
$
there should be non-vanishing at least for a pair of indices $i,j$.

\section{GTBQC for circuit models}\label{sec:ourscheme}

Assume that the server has a circuit-model quantum computer. 
In order to perform BQC on this hardware, we propose to use gate teleportation to implement one-qubit gates.
 The client needs to be able to prepare $\ket{\Phi_{V_i}}$ where the set $\cV=\{V_i\}$ is sufficient to construct universal one-qubit gates. 
 Thus, if the client wants the gate $V_i$ to be performed on a qubit of the server's quantum register, the client first prepares $\ket{\Phi_{V_i}}$ (in which $V_i$ is applied on the second qubit of the Bell pair) and sends its qubits to the server. The client next instructs the server to perform a Bell-basis measurement on the first qubit of the pair and the selected qubit of the register, and consider the second qubit of the pair as the updated qubit of the register (we call such procedure ``GT procedure'').
 This amounts to applying the gate $V_i\sigma,\ \sigma\in\Sigma$ rather than the intended $V_i$ operation. 
The server classically communicates the resultant $\sigma$ to the client.
One can remedy the problem of $V_i \sigma$ rather than $V_i$ through the following adaptive approach.
The client sends a series of $\ket{\Phi_{V^{(j)}}}$  depending on the results of previous measurements.
A desired result is a result of a Bell-basis measurement when $\sigma^{(j)}=\openone$.
At the first step, the client applies $V^{(1)}=V_i$ using the GT idea.
If this action fails, i.e., $V^{(1)}\sigma^{(1)}$ (with $\sigma^{(1)} \neq \openone$) has been implemented instead, the client sends another state $\ket{\Phi_{V^{(2)}}}$ with $V^{(2)}=V_i\sigma^{(1)} {V^{(1)}}^\dagger$.
Continuing on this fashion, if in the first $l$ steps the desired result has not been achieved, for step $l+1$ the client sends $\ket{\Phi_{V^{(l+1)}}}$, where
\begin{equation}
V^{(l+1)}=V_{i}\sigma^{(l)} {V^{(l)}}^\dagger.
\end{equation}
This stochastic process succeeds on average in 4 steps, because each GT has a $1/4$ chance of success independent of the previous outcomes. 
This procedure completes the process of applying single-qubit unitary operators.

It is possible to choose the set $\cV$ and $W$---the two-qubit gate which the server applies---such that the stochasticity of single-qubit operations can be lifted.
This possibility relies on choosing elements of $\cV$ such that they have a commutativity property with respect to $\Sigma$.
We choose the set $\cV$ and $W$ such that they satisfy the following properties
\begin{subequations}\label{eq:commcri}
\begin{align}
 W\Sigma\otimes\Sigma&\subseteq {\Sigma}\otimes{\Sigma} W, \label{eq:comm2qub} \\
\cV{\Sigma} &\subseteq {\Sigma} \cV \label{eq:comm1qub}.
\end{align}
\end{subequations}
The first relation implies that the $\sigma$ by-products can pass through $W$ (possibly with the new set of by-products). The second relation grantees that for any $V\in\cV$ and $\sigma\in \Sigma$, there exists $V'\in\cV$ and $\sigma'\in\Sigma$ such that $\sigma' V = V' \sigma$.
Thus if the client wants to implement $V$, given that the overall by-product of the previous GTs is $\sigma$, the client needs to have the server perform the operation $V'$ through GT.
To design a BQC protocol using the above property, one need to determine suitable $W$, and $\cV$.

For a universal quantum computation, we also need to hide placement of two-qubit gates from the server.
Our approach here is akin to the idea used in Ref. \cite{UBQC-Broadbent-IEEE09}.
Consider the following two-qubit operator:
\begin{equation}\label{eq:2qubitgate}
R(U):=W^\dagger \openone\otimes U W,
\end{equation}
where $U\in\mathrm{U(2)}$.
In general $R(U)$ can be an entangling gate.
\begin{figure}
\centering
  \includegraphics[width=66mm]{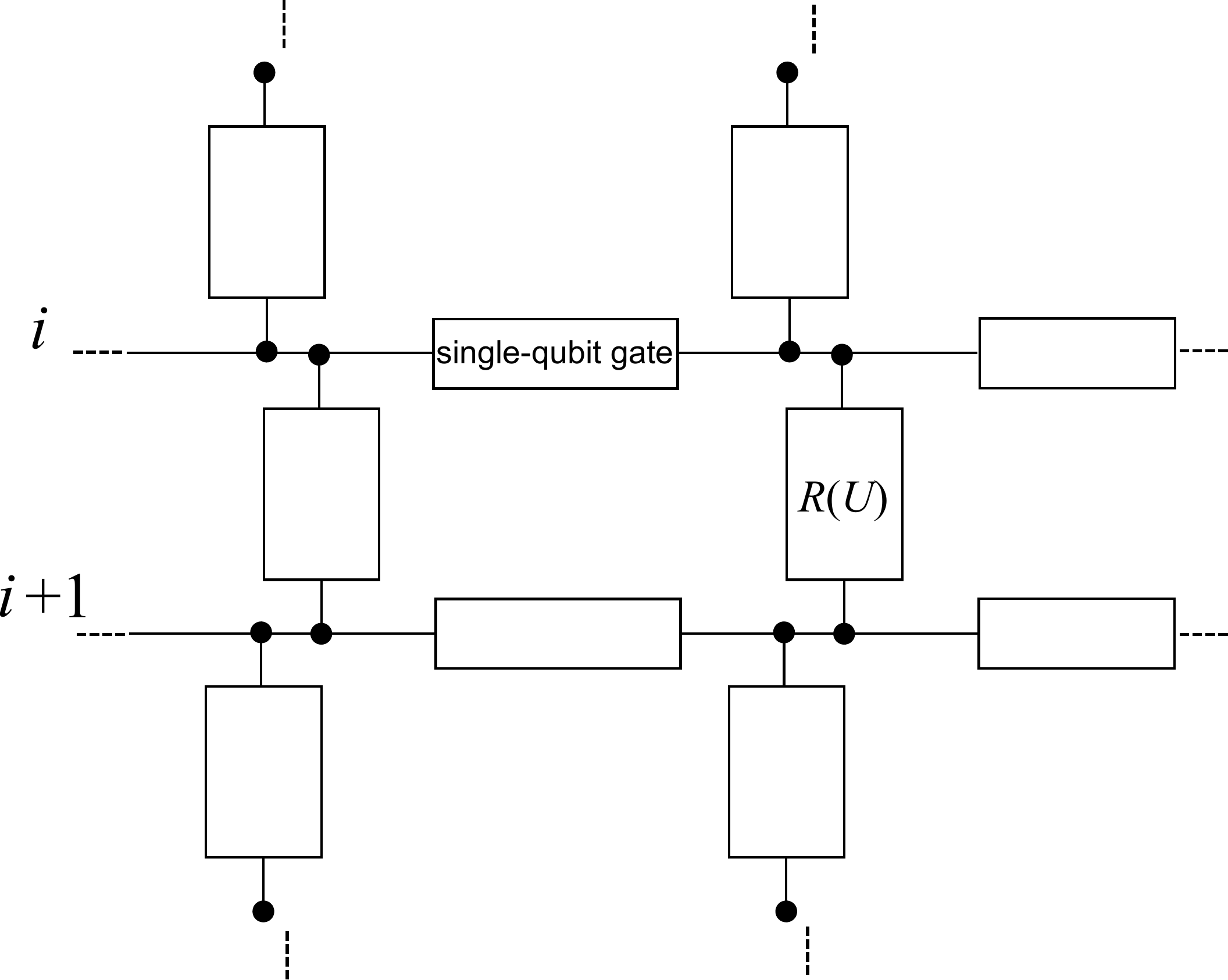}\\
  \caption{Schematic of the circuit-model BQC. The horizontal axes shows the progress of time, while qubits $i$ and $i+1$ (and so on) belong to the server's quantum register. After applying each single-qubit operation between each pair of neighboring qubits some $R(U)$ is also applied.
  This two-qubit operation can be either entangling or identity.}\label{fig:2qubit-spread}
\end{figure}
But the client chooses $U$ either to be the identity or a given operator $S$ for which $R(S)$ is entangling.
To apply $U$ of $R(U)$, the client prepares $\ket{\Phi_{U}}$ and sends it to the server and instructs similar recipe as the single-qubit case.
To hide the placements of the $R(S)$ gates in the computation, the client instructs the server to follow the computation pattern as in Fig.~\ref{fig:2qubit-spread}, which helps spread two-qubit gates (some of which trivial) all over the circuit.

In the following, we first discuss an example of the set $
\cV$ and the two-qubit operation $W$ which satisfy the commutation relations in Eqs.~\eqref{eq:commcri}. 
Next we describe a modified non-stochastic version of our approach where any single-qubit operation in SU(2) is allowed to be performed by both the client and the server. Furthermore, we also discuss a BQC protocol whose privacy does not depend on placing trap qubits.

\subsection{Example of a non-stochastic GTBQC}

We choose the two-qubit gate of the protocol to be the controlled-Z (CZ) gate.
It can be seen that all one-qubit unitaries passing through the CZ gate can be locally written as $Z(\theta) \sigma$ for any $\sigma \in \Sigma$ and $Z(\theta):=\exp(\ii \theta Z/2)$ for any $\theta \in \mathbb{R}$ \cite{PEPSMQC-Koo-PRA14}.
Thus condition~\eqref{eq:comm2qub} is immediately satisfied.

Let us assume 
\begin{equation}
V_i=V_0 Z(\theta_i),
\end{equation}
where $V_0$ is a constant gate.
To satisfy condition~\eqref{eq:comm1qub}, we choose $V_0$ as the member of Clifford group
\begin{equation}
\mathcal{C}_1:=\langle H,\sqrt{Z} \rangle,
\end{equation}
where $H$ is the Hadamard gate.
For example the set $\cV=\{H,HZ(\pm\pi/4)\}$ is a universal set of one-qubit gates. 
One can see that for $V_i \in \cV$, $\{\ket{\Phi_{V_i}}\}$ constitutes a non-orthogonal set.

\subsection{Different example of a non-stochastic GTBQC}
Let the client be able to prepare $\ket{\Phi_{V}}$ for any $V \in$ SU(2).
We also allow that the server to be able to perform all let $\mathrm{SU}(2)$ operators at each step.
We now assume that, in the first step, the client wants to apply the one-qubit gate $U_1$ on a qubit in the server. The client chooses a random $V_1 \in$ SU(2) according to the Haar measure $\mu$ of the group, and sends $\ket{\Phi_{V_1}}$ to the server, which the server should apply the GT procedure on.
If the outcome on the server side is applying $V_1\sigma_1$ (where $\sigma_1$ does not necessarily implies $\sigma_\mathrm{x}$) the client asks the server to perform $R_1=\tau_1 U_1 \sigma_1 V_1^\dagger$ on the same target qubit, where $\tau_1$ has been chosen uniformly randomly from $\Sigma$. Thus, overall the operation $R_1V_1\sigma_1=\tau_1 U_1$ is applied on the target qubit.
By knowing only $R_1$ and $\sigma_1$, the server can not obtain any information about what $U_1$ or $\tau_1$ is, because $R_1$ is as random as $V_1$ (independent of $U_1$ and $\tau_1$), i.e., $\mu(\dd R_1)=\mu(\dd (\tau_1 U_1 \sigma_1 V_1^\dagger))=\mu(\dd V_1^\dagger)=\mu(\dd V_1)$. The role of the random $\tau_1$ in the protocol is to encrypt the state of the server's register qubit by using the quantum Vernam cipher, in order to prevent the server to gain any information about the state of the register. 

Suppose that in the next step, the client wants to perform another one-qubit $U_2$ on the same target qubit. The client chooses a $V_2$ randomly and asks the server to apply it by using GT procedure on the target qubit.
Next the client asks the server to apply $R_2=\tau_2 U_2 \tau_1\sigma_2V_2^\dagger$ on the qubit, where $\sigma_2$ is the by-product of the GT procedure, and $\tau_2$ is chosen randomly from $\Sigma$. Next steps can be applied by following a similar idea.

Here we choose a two-qubit gate $W$ to be the CZ gate.
Now one can choose here an operator $S=X(-\tfrac{\pi}{4})$, where $X(\theta):=\exp(\ii \theta X/2)$, to implement $R(S)$ which is locally equivalent to the CNOT gate.

\subsection{GTBQC without trap}\label{sec:ubqc-without-trap}
In this section, we present a BQC protocol based on the GT idea which does not require frequent checking of the trap qubits to be blind. 
This protocol has hence a privacy independent of the behavior of the server.

Assume that at each step the server can easily implement the set $\cU=\{U_i\}$ of one-qubit gates, whose elements are all necessary for universal quantum computation.
Let $\cV:=\{V_{ij}=U_j^\dagger U_i\}$.
Now suppose that the client wants to apply the gate $U_i$ on a qubit in the server's register. 
Rather than asking the server to apply $U_i$, the client chooses a random $j\in\{1,\dots,|\cU|\}$, sends the state $\ket{\Phi_{V_{ij}}}$ to the server, and ask the server to perform the GT procedure.
In the case that $\sigma=\openone$ the client has the server apply $U_j$ on the register qubit, which means that overall the client has instructed the server to apply $U_j V_{ij} = U_i$ on the qubit without revealing $i$. 
To preclude revealing any information out of $\ket{\Phi_{V_{ij}}}$ we append a secret random $\tau \in \Sigma$ at each step to $U_j V_{ij}$, namely, $U_j V_{ij} \tau$.
This random $\tau$ effectively mixes the sent quantum stat $\ket{\Phi_{V}}$, because we have
\begin{equation}
\sum_{\tau \in \Sigma} \frac{1}{4}\,\ketbra{\Phi_{V\tau}}{\Phi_{V\tau}}=  \frac{\openone}{2} \otimes \frac{\openone}{2},
\end{equation}
for any single-qubit unitary $V$.

It is not necessary for the rest of the discussion that $V$ has the form as $V_{ij}=U_j^\dagger U_i$. 
Assume that the client wants the operation $U\in\cU$ to be performed blindly. Thus $V$ should be applied by the GT procedure, and $U'\in \cU$ must be communicated through the classical channel and next be applied by the server.
Taking into account the by-product of the Bell-basis measurement, the by-products of the previous steps, and the secret $\tau$, the  overall applied operation would become $U' V \sigma$,
where $\sigma \in \Sigma$ is random and unknown to the server (due to the inclusion of the secret $\tau$). 
For the operation $U'V\sigma$ to be equivalent to $U$ (up to a Pauli matrix and/or a phase factor), we need to assume extra criteria similar to Eqs.~\eqref{eq:commcri}. 
In particular, we need that for any $U \in \cU$, and any $\sigma \in \Sigma$, there should exist random choices for $U' \in \cU$, $V \in \cV$, and $\sigma' \in \Sigma$ such that (up to some phase)
\begin{equation}\label{eq:1qubnontrap}
U' V \sigma = \sigma' U.
\end{equation}
Blindness of the protocol necessitates that the probability distribution of $U'$ must be independent of $U$ and $\tau$.
In addition, to encrypt the state of the server's register qubit by using the quantum Vernam cipher, $\sigma'$ should be random and unknown to the server, which is the case if $U'$ is chosen independent of $\tau$.
We also need criterion~\eqref{eq:comm2qub} to hold for the two-qubit gate $W$.

Given all the above criteria, let us see how the client can instruct the server to apply blindly the one-qubit gate $U \in \cU$ on one of its register qubits. 
The steps of this implementation are as follows:
 (i) The client chooses $V$ such that $U'$ of  Eq.~\eqref{eq:1qubnontrap} becomes a member of $\cU$, independent of $U$ and $\sigma$.
 (ii) The client prepares $\ket{\Phi_{V}}$, sends it to the server, and asks the server to apply the GT procedure.
(iii)  The client chooses a random $\tau \in \Sigma$ and finds $U'\in\cU$ from Eq.~\eqref{eq:1qubnontrap} and asks the server to apply $U'$ (communicated classically to the server).

A specific example of the operations meeting the criteria above is given here.
Choose the two-qubit gate of the protocol to be the CZ gate, and let $U_i = H Z(\theta_i)$, where $\theta_i = \pi\, i / 4$ for $0\leqslant i < 8$. 
It is evident that the set $\cU=\{U_i\}$ is a universal set for building any one-qubit gate. 
For $\cV$, choose $\cV=\{V_{ij}|V_{ij}=U_j^\dagger U_i = Z(\theta_i - \theta_j)=Z(\theta_{i-j})\}$. 
One can see that for $\sigma=Z$, $U_j V_{ij} Z = X U_i$, whence choosing a random $j$ makes the server blind about $i$. 
For the case $\sigma=X$ or $Y$, $V_{ij}\sigma = Z(\theta_{i-j})\sigma = \sigma Z(\theta_{j-i}) = \sigma Z(\theta_{i-(2i-j)}) = \sigma V_{i,2i-j}$.  Also $ U_k X = Z U_{-k}$ and $U_k Y = -Y U_{-k}$, where $-k \equiv 8-k$.
Thus, here we have $ U_{j-2i} V_{ij} X = Z U_i$ and $ U_{j-2i} V_{ij} Y = - Y U_i$. 
Note that if $j$ is chosen randomly, this makes $j-2i$ random as well and independent of $i$ and $\sigma$.

\section{GTBQC for MQC models}
In measurement-based quantum computation (MQC), rather than performing unitary gates, quantum computation is implemented by applying a series of adaptive single-site measurements on a highly entangled state \cite{OneWay-Rauss-PRL01,MQC-Nature09}.
Let $\mathcal{M}=\{M_i\}$ be the set of all single-site measurements $M_i$ needed to be performed in an MQC model.
Our idea here is to construct $\cV$ such that its members permute elements of $\mathcal{M}$.
Let us restrict ourselves to projective measurements, where any measurement $M_i$ can be described by a set of orthogonal basis $M_i\leftrightarrow\{\ket{m_{ik}}\}$ (measurement basis).
The unitary operator
\begin{equation}
V_{ij}=\sum_k\ketbra{m_{jk}}{m_{ik}},
\end{equation}
substitutes measurement $M_i$ for $M_j$, if applied to the target qubit before the measurement.
Hence to perform $M_i$ blindly, the client first applies $V_{ij}$ via the stochastic GT procedure introduced in Sec.~\ref{sec:ourscheme}, with a random $j$.
Next, the client asks the server to measure the target qubit in the $M_j$ basis.

This process is stochastic works for all MQC models, but requires the client to be able to prepare a relatively large set of entangled states.
We now show how to design a blind MQC model which lifts the need for stochasticity.
In other words, we set up a series of constrains similar to Eqs.~\eqref{eq:commcri} for blind MQC models.
To this end, we use the ``correlation-space MQC'' framework \cite{NovelMQC-Gross-PRL07, MQCbeyondOneWay-Gross-PRA07,PEPSMQC-Koo-PRA14} for its generality (in the sense that this framework can describe all known MQC models).

\subsection{GTBQC for correlation-space MQC}
We first briefly review the ideas of the correlation-space MQC~\cite{NovelMQC-Gross-PRL07, MQCbeyondOneWay-Gross-PRA07,PEPSMQC-Koo-PRA14}.
A matrix product state (MPS) is a state that can be written in the standard basis as follows:
\begin{multline}\label{eq:mps}
\ket{\mathrm{MPS}} = \\
\sum_{i_1,i_2,\dotsc,i_N} \bra{L} A_N(i_N)\dotsm A_2(i_2)A_1(i_1) \ket{R}\:
\ket{i_N,\dotsc,i_2,i_1},
\end{multline}
where $N$ is the number of sites, $0 \leq i_j < D$, and $A_j(i)$ is a $d\times d$ matrix attributed to
 state $\ket{i}$ of site $j$. 
Here $\bra{L}$ and $\ket{R}$ are the left and the right boundary vectors. 
The amplitude coefficients of the MPS are computed by a set of matrix multiplications in the so-called \emph{correlation space}.
 If we measure the $j$th site and the output state is $\ket{\varphi}$, then the MPS becomes
 \begin{multline}\label{eq:mps_phi}
  \ket{\varphi}_j\bra{\varphi} \,
 \ket{\mathrm{MPS}}\; =\\ \sum_{i_1,\dotsc,i_{j-1},i_{j+1},\dotsc,i_N}
 \bra{L} A_N(i_N)\dotsm A_j[\varphi] \dotsm A_1(i_1) \ket{R}\\
 \ket{i_N,\dotsc ,i_{j+1}}\ket{\varphi}\ket{i_{j-1},\dotsc ,i_1},
 \end{multline}
 where $A_j[\varphi]:= \sum_i\varphi^\ast_i A_j(i)$ and  $\ket{\varphi}=\sum_i \varphi_i\ket{i}$. 
Hence, any operator in $\Span( A_j ):= \{\sum_i \alpha_i A_j(i)\}$ can be realized at the correlation space by a single-site measurement. If we measure, e.g., $l$ sequential sites after site $m$ of the MPS, we obtain the following operator in the correlation space:
 \begin{equation} \label{eq:prod-U}
 U=A_{m+l}[\varphi_l]\dotsm A_{m+2}[\varphi_2]A_{m+1}[\varphi_1].
 \end{equation}
  Thus, by choosing appropriate $A_j$ as the lists of matrices and suitable measurement bases,
 one can construct any unitary in U($d$) at the correlation space. 
 Let us select $A_j$ and measurement bases such that $A_j[\varphi]$ for any measurement outcome at site $j$ is a unitary operator.
Randomness of measurement outcomes causes that instead of a desired $\ket{\varphi}$ as outcome one obtains $\ket{\varphi'}$. Therefore, rather than $U:=A_j[\varphi]$, $A_j[\varphi']$ is realized in the correlation space. In such cases, we consider $A_j[\varphi']$ as
  \begin{equation}\label{eq:err}
 A_j[\varphi'] = E\,U,
 \end{equation}
 where $E=A_j[\varphi'] U^\dagger$ is the so-called ``by-product operator.''
 Such by-products need to be somehow circumvented. 
 Later, in Eq.~\eqref{eq:mangbypr}, we will present conditions to manage by-products.
 
 Let us assume translationally-invariant correlation-space MQC model, where $A_j=A$. 
 The model is assumed to be deterministic, that is, the number of steps needed to perform any gate is predetermined.
 Let $\mathcal{M} =\{M_i\}$ be the set of all single-site measurement $M_i$ of the MQC model, which are used to implement one-qubit gates.
 Let $\cU$ be the set of all single-qubit unitaries $U_i$ that can be performed at each step of the MQC model, which can be taken as $A[\varphi_i^0]$, where $\ket{\varphi_i^0}$ is the target measurement result of $M_i$.
 Let $\cE$ denote the set of all manageable by-products of the  MQC model. Thus, to manage by-products, for any $U_j\in\cU$ and any $E \in \cE$, there should exist an $M_i$ such that for any $\ket{\varphi_i}$ in $M_i$, we have
  \begin{subequations}\label{eq:mangbypr}
  \begin{align}
  A[\varphi_i]E \,U_j^\dagger &\in\cE, \label{eq:byprodef} \\
 W\cE\otimes\cE &\subseteq \cE\otimes\cE W, \label{eq:byprodcommmqc}
  \end{align}
  \end{subequations}
where $W$ is the two-qudit gate implemented in the MQC model.
Let $\cV$ be the set of one-qubit unitaries from which  $\Phi_V$ is constructed.
When the client wants the gate $U_i$ to be implemented by the server, the sufficient criterion for our non-stochastic blind MQC model to work is that for any $\ket{\varphi_i}\in M_i$ and any $\sigma\in\Sigma$, there should exist an $M_j$ and $V \in \cV$ such that
\begin{equation} \label{eq:mqccrit}
 A[\sigma V^\dagger \ket{\varphi_j}] A^\dagger[\varphi_i] \in \cE,
\end{equation}
 for any $\ket{\varphi_j}\in M_j$.
 Hence, to blindly perform an $A[\varphi_i]$ on the server's register site, the client sends $\ket{\Phi_V}$ and asks the server to perform the GT procedure, next asks the server to measure the target qubit in the $M_j$ basis.
\subsection{Example of a GT-based blind MQC model}
One way to satisfy criterion~\eqref{eq:mqccrit} is to have
$\sigma V^\dagger \ket{\varphi_j} \in M_i$, for some $M_i$.
Here we explicitly construct such set $\cV$ suitable for the one-way MQC model \cite{OneWay-Rauss-PRL01}.
In this model, the measurement basis is parametrized as
\begin{equation}
M_\theta = \{\ket{\theta\pm}:=(\ket{0}\pm\e^{\ii \theta}\ket{1})/\sqrt{2}\}.
\end{equation}
Now consider
\begin{equation}
V_\gamma := Z(-\gamma)=\exp (-\ii \gamma Z /2).
\end{equation}
By applying this operator for a random $\gamma$ through the GT procedure before a measurement $M_\beta$ (with $\beta=\alpha-\gamma$), the server will have $M_\alpha = M_{\beta+\gamma}$ applied on his qubit.
Hence, in this manner, the client can hide $\alpha$ from the server.
Note that here the application of the Pauli matrices is also a well behaving operation because
\begin{align}
M_\alpha &\xrightarrow{X} M_{-\alpha}, \\
M_\alpha &\xrightarrow{Z} M_{\alpha+\pi}.
\end{align}
Thus by choosing $\beta$ in accordance with the above conditions (and depending on the output of the Bell-basis measurement) criterion~\eqref{eq:mqccrit} will be satisfied.
For a discrete example, one can choose legitimate $\theta$s for $M_\theta$ from  $\{\pi, \pi/2, \pm \pi/4\}$. 
\section{Concluding remarks}
We have shown that using the idea of storing one-qubit gates in entangled pairs (``the gate-teleportation procedure''), a client can perform blind quantum computation on a remote server.
We have demonstrated our idea for quantum circuit model and  measurement-based quantum computation.
We have illustrated our proposition for blind quantum computation through four specific examples.

One remark is in order here. An alternative method for the client to perform one-qubit gates, rather than storing in an entangled pair and having the server perform a gate-teleportation operation (as we discussed earlier), is to receive qubits from the server's register and apply the one-qubit gates on her own side and send the resultant qubit to the server. 
However, our proposed gate-teleportation approach has several advantages over this alternative scenario.
(i) In the qubit sending method, the quantum channel needs to have the extra property of preserving possible entanglement of the sent qubit with the rest of the server's register.
In contrast, since in our method the entangled pairs sent from the client to the server are carried locally (with no residual entanglement with any other qubits) no such long-distance entanglement preservation property is required.
(ii) In the non-stochastic version of our approach, the client can in principle send all the computation-carrying entangled pairs (with appropriate sequence) once to the server.
However in the alternative approach sending and receiving register qubits are needed in order to apply each single-qubit operation.
 Thus our approach has a relative speedup with respect ot the alternative qubit-sending method.
One can also apply the similar idea of sending all entangled qubits once even for the stochastic version of our method, possibly with some extra overhead.
(iii) In our approach it is straightforward to calculate the probability of detecting a privacy-invasive/non-cooperative server by the trap method, because the set $\cV$ is independence of the server's quantum register. 
However, in the alternative approach, since the client does not necessarily know the state of the register qubit sent from the server, it seems difficult how to estimate the probability of revealing a non-cooperative server.
(iv) In some cases it may not be practically possible to transfer any qubit of the server's quantum register.

Some interesting line to pursue for future progress, for example, are to see how one can use multipartite entanglement for blind quantum computation, investigate its potential benefits, an see how possibly one can reduce necessary resources for this sort of computation, or improve its overall performance. 

\section*{Acknowledgments}
We would like to thank A. T. Rezakhani and V. Karimipour for useful comments. 
We would also like to acknowledge J. Fitsimons and M. Tomoya for clarifications on the privacy of the protocol described in Ref.~\cite{UBQC-Broadbent-IEEE09}.

\bibliographystyle{apsrev4-1}
\bibliography{refs}

\end{document}